# BUDD: Multi-modal Bayesian Updating Deforestation Detections

*Alice M.S. Durieux[†,1], Christopher X. Ren[†,2], Matthew T. Calef[1], Rick Chartrand[1], Michael S. Warren[1]*

[1]Descartes Labs, Inc, 100 N Guadalupe St, Santa Fe, NM
[2]Intelligence and Space Research Division, Los Alamos National Laboratory, Los Alamos, NM
† These authors contributed equally.

## ABSTRACT

The global phenomenon of forest degradation is a pressing issue with severe implications for climate stability and biodiversity protection. In this work we generate Bayesian updating deforestation detection (BUDD) algorithms by incorporating Sentinel-1 backscatter and interferometric coherence with Sentinel-2 normalized vegetation index data. We show that the algorithm provides good performance in validation AOIs. We compare the effectiveness of different combinations of the three data modalities as inputs into the BUDD algorithm and compare against existing benchmarks based on optical imagery.

## 1. INTRODUCTION

Global forest loss is a major concern with severe consequences including atmospheric carbon accumulation and biodiversity reduction [1]. In order to enable preventive actions in the field, remote-sensing based forest monitoring systems must provide accurate and timely deforestation alerts. The highest rate of deforestation is currently occurring in tropical areas [2], which are also the cloudiest. For instance, in the Brazilian state of Amazonas, only 15% (3610 scenes) of the 22623 Sentinel-2 images acquired in 2019 had a cloud fraction below 10%. The scarcity of cloud free imagery makes it challenging for existing forest monitoring tools, which largely rely on optical imagery [1], to provide timely deforestation alerts in the tropics. Synthetic Aperture Radar (SAR) can image the surface of the Earth regardless of cloud cover by emitting electromagnetic radiation and quantifying the amplitude of the reflected signal (backscatter), or leveraging the phase information contained in the reflected SAR signal to quantify to phase similarity between pairs of SAR images (InSAR coherence). Both backscatter and coherence can be used to distinguish forest from bare soil; dense vegetation reflects a large portion of the signal (high backscatter), but small movements in the leaves and stems between SAR collections result in low coherence between consecutive images. Bare earth has the opposite effect on backscatter and coherence respectively, thus making SAR a potentially effective modality to detect deforestation and enhance existing optical-based alerts.

In this study, we adapt a probabilistic approach that allows for the integration of multiple remote sensing modalities [3] to monitor forest cover in the Brazilian states of Amazonas and Mato Grosso (MG), from 2018/01/01 to 2019/12/31. This work presents a novel approach in leveraging coherence data with SAR backscatter and normalized difference vegetation index (NDVI) to develop a probabilistic model for deforestation alerts.

## 2. METHODS

### 2.1. Data acquisition and processing

This study was completed using the Descartes Labs platform [3], which stores publicly available satellite imagery including Sentinel-1 (C-band SAR sensor launched in 2014) and Sentinel-2 (multispectral instrument launched in 2015) in a cloud remote object storage. We analyzed all Sentinel-1 and Sentinel-2 images acquired between 01/01/2015 and 12/31/2019 in test areas of interest (AOIs) selected over Mato Grosso and Amazonas. The AOIs were decomposed into approximately 1200 tiles of 512 x 512 pixels at 20m resolution to allow for parallel processing of the data. Sentinel-2 scenes with a cloud fraction larger than 15% were excluded, and clouds were masked using Descartes Labs' proprietary dlcloud model prior to the computation of NDVI.

Sentinel-1 data were split by pass (ascending or descending) and relative orbit. We computed the VV/VH band ratio, as well as interferometric coherence between pairs of consecutive images of the same orbital track [4]. The NDVI, VV/VH ratio and coherence time series were co-registered and a spatiotemporal denoising algorithm based on total variation regularization [5] was applied to reduce speckle. Examples of each data modality taken in the Amazonas are shown in Figure 1.

To identify forested regions, we used a benchmark land cover dataset published by the Brazilian National Institute for Space Research in 2017 [6]. Seasonality in the time series was detrended by subtracting the $90^{th}$ percentile of the distribution of forested pixels (as defined by the land cover dataset) within each 512 x 512 tile for each pixel [7].

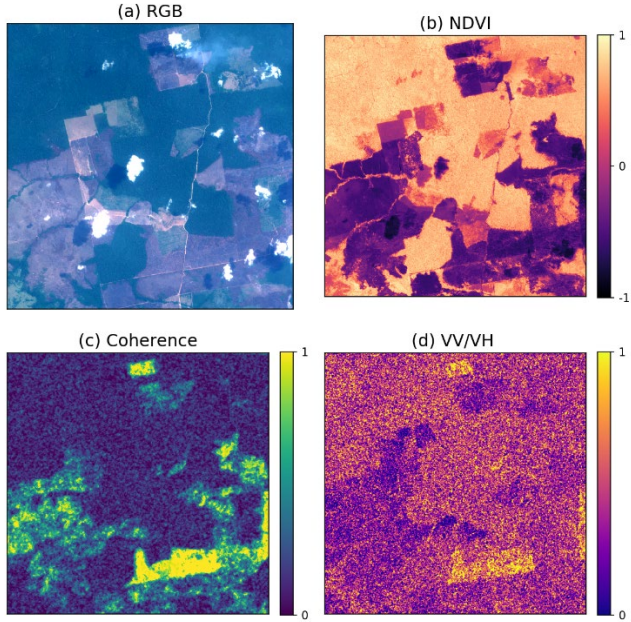

**Figure 1:** Example multi-modal dataset: (a) RGB (b) NDVI (c) Coherence (d) VV/VH Ratio

### 2.2. Bayesian Updating Deforestation Detections (BUDD)

The Bayesian Updating Deforestation Detection (BUDD) algorithms shown in this work leverage a pixel-wise probabilistic algorithm fully described in [3]. We split the NDVI, VV/VH and coherence time series into a forest defining period (from 2015/01/01 to 2017/12/31), and a forest monitoring period (2018/01/01 to 2019/12/31).

For each pixel, we used the median and standard deviation of all observations in the forest defining period to parametrize a Gaussian distribution of the forest signal. BUDD leverages the property that forested pixels exhibit high NDVI, high backscatter and low coherence while bare earth displays the opposite pattern, to derive the non-forest distributions for each pixel by shifting the mean of the forest Gaussians for each sensor by a set number of standard deviations, chosen empirically (-6, -6 and 7 for NDVI, VV/VH and coherence respectively). This allows for ease of deployment and transferability between different types of forest as one only needs to estimate the distance between the means of the forested and non-forested distributions for a given area in order to deploy BUDD.

The probability density functions (PDFs) of the forest and non-forest distributions are then used to compute the probabilities of a new observation belonging to either distribution in the monitoring time series. From these probabilities, the conditional probability of the pixel belonging to the non-forest distribution is derived by applying Bayes theorem. Observations acquired on the same day were combined into a joint probability of these observations belonging to the non-forest distribution. Potential deforestation events were flagged when the conditional probability of forest loss given an observation exceeded 0.6. Flagged pixels were entered into a continuously updating Bayesian computation to update the probability of forest loss using subsequent observations. This was continued until the probability of forest loss either exceeded 0.975, in which case the pixel was marked as deforested; or dropped below 0.5, in which case the flag was removed and normal monitoring resumed. At least two observations were required to classify a pixel as deforested.

### 3. RESULTS

#### 3.1. Input datasets

Summary statistics on the images included in the forest defining period are available in Table 1. Few NDVI images were available due to the stringent cloud fraction requirement.

**Table 1:** Summary statistics of the number of images included in the forest defining period (2015/01/01 to 2017/12/31).

|  |  | Amazonas | Mato Grosso |
|---|---|---|---|
| **NDVI** | Mean (std) | 19 (14) | 38 (23) |
|  | Range | 2 - 177 | 11 - 227 |
| **VV/VH** | Mean (std) | 114 (84) | 83 (33) |
|  | Range | 29 - 443 | 38 - 244 |
| **Coherence** | Mean (std) | 50 (24) | 47 (25) |
|  | Range | 4 - 203 | 4 -143 |

#### 3.2 Detection Comparison

Here we show deforestation detections over the AOI displayed in Figure 2.a (Sentinel-2 image acquired on 2017/09/03) and 2.b (Sentinel-2 image acquired on 2019/11/03). Figures 2.c, 2.d, and 2.e are the deforestation detections from the NB, BC and NBC-BUDD models respectively. Figure 2.f shows a direct comparison between our best performing algorithm (NBC) and the GLAD deforestation alerts, the Landsat-based deforestation monitoring gold standard [1].

Figures 2.c and 2.d show that including coherence in the BUDD algorithm generally increases the coverage of the deforestation detections, as shown by the greater coverage green detections. This is likely due to the fact that the addition of coherence data allows the BUDD algorithm to detect 'coherent' transitions over areas from forest to bare-earth following forest loss as coherence is calculated with a spatially windowed average.

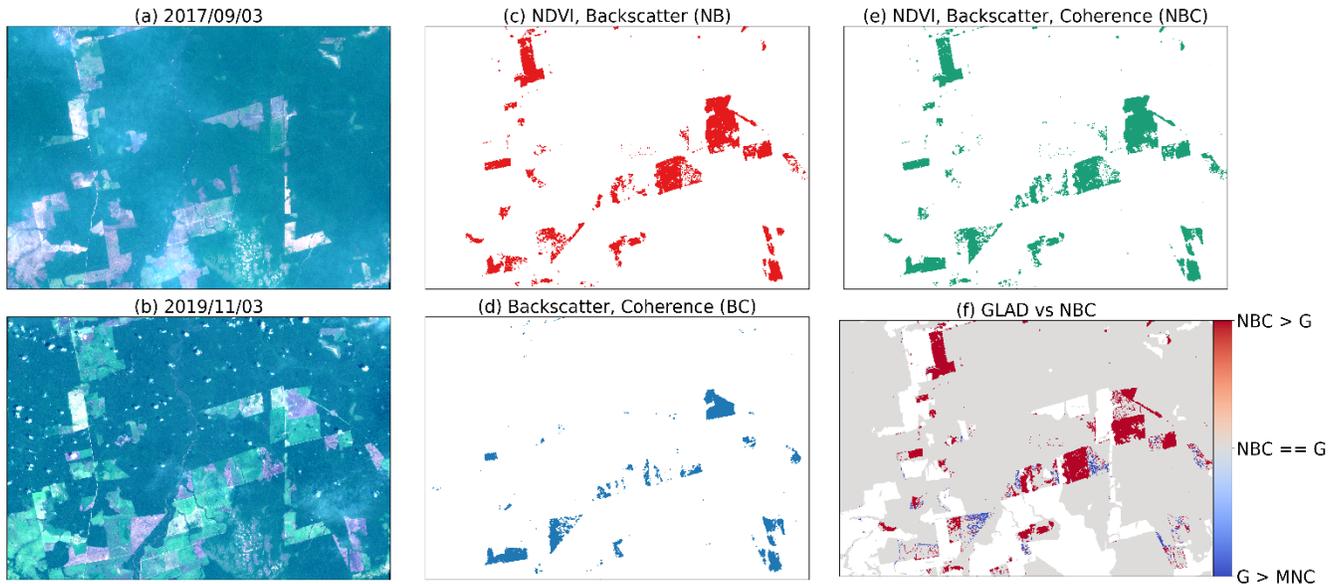

**Figure 2** a) 'before' (2017/09/03) and b) 'after' (2019/09/03) optical image c) NB, d) BC and e) NBC deforestation detections over a small, representative AOI chosen in the test AOI of Amazonas. (f) shows a quantitative comparison between the NBC detections and the GLAD (G) detections.

Figure 2.d shows the results of a BUDD model with only SAR backscatter and coherence (BC). We note that removing NDVI as an input for the model results in an increased rate of false negative detections. We hypothesize that the BC-BUDD algorithm detects deforestation over areas that have been razed/burnt with little regrowth/remaining vegetation, resulting in a greater signal in the backscatter and coherence, as evidenced by the overlap between the BC-BUDD detections and the brown areas in Figure 2.b.

Finally, we show a direct comparison between GLAD alerts and NBC in Figure 2.f. In red we show areas detected as deforested by NBC but not by GLAD, in grey where the NBC and GLAD alerts agree (either as detections or non-detections) and in blue detections reported by GLAD but not by NBC. We note that the majority of the red pixels and blue shown here are likely to be true positives, as evidenced by examining Figure 2.b. As GLAD relies on optical data to perform detections, we hypothesize that the lack of cloud free data over the red areas hinders the performance of the algorithm over these areas.

### 3.3 Detection Statistics over Mato Grosso and Amazon

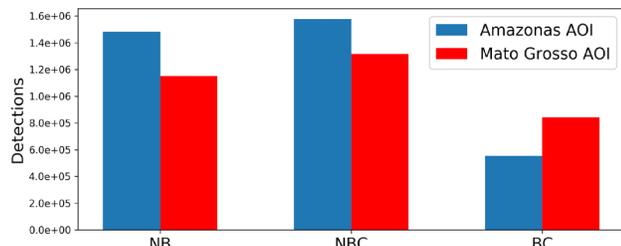

**Figure 3:** Total detections for the NB, NBC and BC-BUDD method over Amazonas and Mato Grosso AOIs.

Figure 3 shows the total number of detections collected over the Amazonas and Mato Grosso test AOIs respectively. We note that the NBC-BUDD method returns more detections than both NB and BC methods. Based on our visual validation of the detections performed over the test areas (some of which is shown in Figures 2, 4 and 5), the majority of the additional detections for the NBC-BUDD algorithm are likely to be true positives. However, further work is needed to confirm this over larger areas.

## 4. DISCUSSION

Figure 3 shows how the spatial averaging window used to calculate coherence improves detection coverage over affected areas. We also note the benefit of including NDVI as an input for BUDD in Figure 4: Figure 4.c demonstrates the capability of using only SAR data inputs (BC) in detecting areas that have been clearly burnt/razed to bare earth. However, areas where some vegetation remains following deforestation result in false negatives. Including NDVI in the BUDD inputs helps alleviate this issue as shown in Figure 4.d. We note that backscatter is related to vegetation density, as the microwaves scattered off vegetation increase with increasing density. The fact that NBC detections perform better than BC over the areas shown in Figure 2 suggests that the distributions of forest and non-forest NDVI values are more separable than those of the forest and non-forest backscatter and coherence distributions.

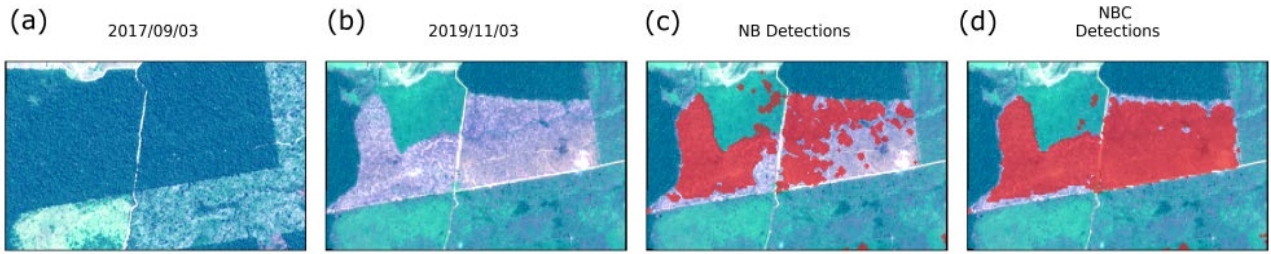

**Figure 4:** The effectiveness of coherence: (a) before (b) after deforestation Sentinel-2 images (c) NB detections (d) NBC detections.

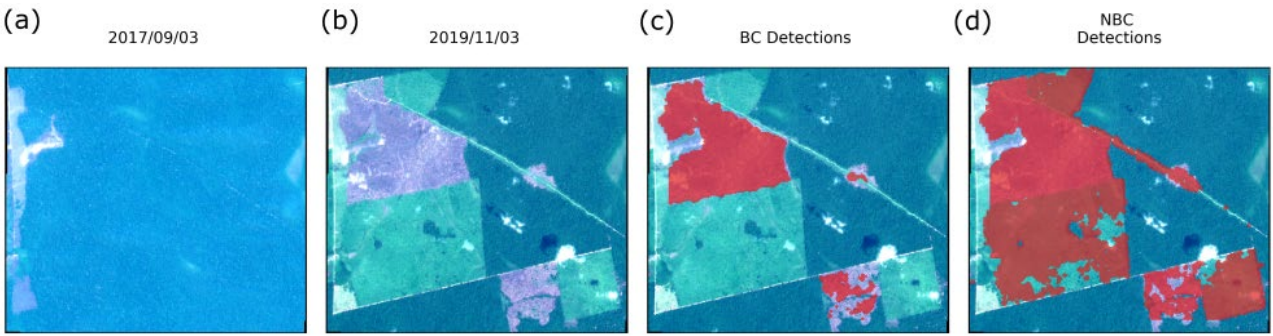

**Figure 5:** The effectiveness of NDVI: (a) before, (b) after deforestation Sentinel-2 images, (c) BC detections, (d) NBC detections.

One of the main issues in detecting deforestation in cloudy areas such as the Amazon is the lack of accurate ground truth. The development of a more accurate forest mask may help mitigate false negative detections. Similarly, we have observed that suboptimal cloud masking can distort the estimation of the forest and non-forest distributions and lead to detection errors. Nonetheless, deriving forest and non-forest distributions using the proposed pixel-based approach allows for the incorporation of pixel-specific variability in the input data modalities, a major advantage over approaches that use forested and non-forested defining data representing the entire study area. Additionally, Figure 2.f demonstrates that the incorporation of SAR data can greatly improve the rate of detection of deforestation in cloudy areas. However, further validation of the comparison between NBC and GLAD alerts is necessary over larger areas to quantify the abilities of both detection methods. Future work will include comparing the timeliness of both methods in determining that a pixel was deforested.

In conclusion, we have shown that the incorporation of InSAR coherence in the BUDD algorithm enhances the performance of deforestation detection by providing better surface coverage. We have also shown that the NBC-BUDD algorithm is likely the most effective in detecting deforestation, as the information contained in multiple imaging modalities allows for a more accurate modelling of forest and non-forest distributions.